\documentclass[12pt,a4paper]{article}
\usepackage{a4wide}
\usepackage{amsfonts}
\begin{document}
{\renewcommand{\thefootnote}{\fnsymbol{footnote}}
\hfill  PITHA -- 00/19\\
\medskip
\hfill gr--qc/0008054\\
\medskip
\begin{center}
{\LARGE  Angular Momentum in Loop Quantum Gravity}\\
\vspace{1.5em}
M.~Bojowald\footnote{e-mail address: 
{\tt bojowald@physik.rwth-aachen.de}}\footnote{New address: 
Center for Gravitational
    Physics and Geometry, Department of Physics, The Pennsylvania
    State University, University Park, PA 16802, USA}\\
Institute for Theoretical Physics, RWTH Aachen\\
D--52056 Aachen, Germany\\
\vspace{1.5em}
\end{center}
}

\setcounter{footnote}{0}

\newcommand{\half}{{\textstyle\frac{1}{2}}}
\newcommand{\lP}{l_{{\mathrm{P}}}}
\newcommand{\mod}{\mathop{{\mathrm{mod}}}}
\newcommand{\md}{{\mathrm{d}}}
\newcommand*{\N}{{\mathbb N}}
\newcommand*{\Z}{{\mathbb Z}}

\begin{abstract}
  An angular momentum operator in loop quantum gravity is defined
  using spherically symmetric states as a non-rotating reference
  system. It can be diagonalized simultaneously with the area operator
  and has the familiar spectrum. The operator indicates how the
  quantum geometry of non-rotating isolated horizons can be
  generalized to rotating ones and how the recent computations of
  black hole entropy can be extended to rotating black holes.
\end{abstract}

\section{Introduction}

In General Relativity familiar observables like energy or (angular)
momentum, which are related to space-time symmetries, can be defined
only in special regimes because in the general situation there is no
reference frame with respect to which those symmetries could be
defined. The usual procedure then is to introduce boundaries which are
endowed with additional structure determined by suitable boundary
conditions. Using the structure at the boundary, observables can be
defined as functionals of the boundary values of the gravitational
field.

As an example, recall the situation of angular momentum defined at
spatial infinity of a space-time. Classically, one has to fix an
asymptotic Cartesian frame carrying an $SO(3)$-symmetry with respect
to which angular momentum can be defined for an asymptotically flat
space-time satisfying appropriate fall-off conditions. One can then
read off the components of angular momentum by comparing the
asymptotically flat metric with the fixed flat one or, in a
Hamiltonian formulation, construct generators of rotations along the
Killing vector fields of the flat metric \cite{RT}. In either way, one
needs to fix a reference frame which is regarded as being
non-rotating, and to specify fall-off conditions for asymptotically
flat metrics, which build the subclass of space-times for which
angular momentum at spatial infinity can be defined.

Asymptotic boundary conditions for gravity in Ashtekar's formulation
have been discussed in Ref.\ \cite{AsFlat}.  In a connection
formulation with its internal $SU(2)$-gauge group, boundary conditions
at spatial infinity not only involve specifying the fall-off of
dynamical fields but also fixing an internal direction in
$SU(2)$-space because gauge transformations are frozen at spatial
infinity. In polar coordinates $(r,\vartheta,\varphi)$ of the fixed
reference frame of a non-compact space manifold $\Sigma$, the fall-off
conditions for the connection state that all its components fall off
at least with an order of $r^{-2}$ at spatial infinity. Fixing the
internal direction of the radial component $A_r^i$, which is the only
one we need for the present purpose, is more crucial because, as we
will see, it relates the internal spin to angular momentum. (At first
sight it may seem alarming to relate angular momentum to operations in
an internal gauge group, even more so because this is possible only if
the internal group coincides with the group $SU(2)$ of rotations which
could be a mere coincidence. Note, however, that in our case the
internal gauge group of gravity is the local group of dreibein
rotations which at the boundary are fixed and tied to global rotations
by the phase space structure. In this context it is quite natural to
have a relation between internal spins at the boundary and angular
momentum.) The simplest choice may be to choose a constant internal
direction (independent of $\vartheta$, $\varphi$), but this is
inappropriate for an asymptotically flat connection which, as
discussed in Ref.\ \cite{AsFlat}, should have odd parity of its
leading order term. We thus are lead to an asymptotic radial component
of the connection having the form
\begin{equation}\label{Asympt}
  A_r^i=r^{-2}A\,n_r^i(\vartheta,\varphi)+O(r^{-3})
\end{equation}
where $A$ is independent of the polar coordinates and
$n_r:=(\sin\vartheta\cos\varphi, \sin\vartheta\sin\varphi,
\cos\vartheta)$.

In light of our discussion above it is also worth mentioning that one
can reverse the argumentation: starting from a boundary which
topologically is a two-sphere but carries no additional structure, a
suitable fixed internal direction (which should have mapping degree
one) provides a bijective map $\partial\Sigma\to S^2$ which can be
used to endow the boundary with an $SO(3)$-action and thereby with a
reference frame with respect to which angular momentum is to be defined.
This view demonstrates why the fixed internal direction will play a
prominent role in our definition of an angular momentum operator.

We next have to find a quantum formulation of the asymptotic structure
by giving conditions for states to be considered as being
asymptotically flat. Recalling that the kinematical Hilbert space of
loop quantum gravity is a space of (cylindrical) functions on the
space of generalized connections, the fall-off conditions can be
imposed by constraining the support of an asymptotically flat state to
only asymptotically flat connections. This immediately leads to the
tangle property, which has been assumed in Ref.\ \cite{QSDVI} in order
to quantize the ADM energy, of asymptotically flat spin network
states: an allowed spin network state has only transversal edges
intersecting the boundary at infinity; for a family of holonomies
$h_r$ each lying in an orbit at radius $r$ converges to unity for
$r\to\infty$ because of the $r^{-2}$ fall-off of the connection
components, and a spin network state cannot depend non-trivially on
this trivial holonomy.

It is, however, not immediate to see how the reference frame, with
respect to which angular momentum will be defined, is realized in the
spin network quantization. In order to illustrate its role we will
first discuss possible standard approaches to a quantization of
angular momentum. First, one can try to start from a classical
expression of angular momentum, e.g.\ that of Ref.\ \cite{AsFlat}
\[
  L_{\mathrm{ADM}}[N^a]=\int_{\partial\Sigma}\md^2y\,
  n_aN^{[b}E_i^{a]}K_b^i
\]
associated with an asymptotic rotation generated by the vector field
$N^a$, and then follow standard quantization steps. However, this
expression contains the extrinsic curvature which would be quantized
to a commutator of the volume operator with the Euclidean Hamiltonian
constraint \cite{QSDI} resulting in a very complicated operator.

A second procedure could be to start from the simple action of
rotations on spin networks (simply rotating the graph; this is the
usual action of the diffeomorphism group where, however, rotations are
not included in the gauge group of diffeomorphisms because those have
to be asymptotically trivial) and to determine angular momentum as
their generators by differentiation. However, this does not work
because as with the diffeomorphism constraint the action used on the
space $\Phi_{\Sigma}$ of cylindrical functions is not strongly
continuous and so its generators do not exist \cite[Appendix
C]{ALMMT}.

\section{The Angular Momentum Operator}

Our proposal to remedy this situation is to use spherically symmetric
states \cite{SymmRed} which are distributional states in the
kinematical Hilbert space of loop quantum gravity being exactly
symmetric with respect to a given action of the rotation group
(classified by a conjugacy class $[\lambda]$ of homomorphisms from the
$U(1)$-isotropy group of a general point in $\Sigma$ to the gauge
group $SU(2)$). All those symmetric states, which are defined as
distributions being supported on the space of rotationally invariant
connections, can be identified with cylindrical functions of connections
and scalar fields on a radial manifold $B$. Roughly, this comes from
the decomposition (up to gauge transformations)
\begin{eqnarray*}
 A(r,\vartheta,\varphi) & = & A_1(r)n_r^i\tau_i\,\md r 
  +2^{-\frac{1}{2}}(A_2(r)n_{\vartheta}^i+(A_3(r)-
  \sqrt{2}\,)n_{\varphi}^i) \tau_i\,\md\vartheta\\
  & &
  +2^{-\frac{1}{2}}(A_2(r)n_{\varphi}^i-(A_3(r)-\sqrt{2}\,)
  n_{\vartheta}^i)\tau_i\,\sin\vartheta\,\md\varphi
\end{eqnarray*}
of an invariant connection into a reduced connection given by $A_1$
and scalar fields $A_2$, $A_3$; for details we refer to Ref.\ 
\cite{SymmRed}.  Vice versa, any cylindrical state in the full theory
can be mapped (``averaged'') to a distributional state in the reduced
formulation giving rise to a map $\rho_{[\lambda]}\colon
\Phi_{\Sigma}\to\Phi'_B$ with $\Phi'_B$ denoting the topological dual
to the space of cylindrical functions on the space of generalized
connections and scalar fields in $B$. This map is defined by viewing a
given cylindrical function $f\in\Phi_{\Sigma}$ as a function defined
only on the subspace of invariant connections, and can be considered
as a pull back of cylindrical functions to the space $\Phi'_B$.
Employing $\rho_{[\lambda]}$ and the pull back to $\Phi'_B$ of general
states serves two purposes: On the one hand, evaluated in an invariant
connection all holonomies to rotated asymptotic edges, which are
transversal to the orbits owing to the tangle property, are gauge
equivalent because connections in the support of the pull backs are
invariant under rotations up to gauge. This implies that rotated spin
networks are gauge equivalent rather than orthonormal, and so
infinitesimal generators of rotations exist being related to
generators of internal rotations by the fixed asymptotic internal
directions. On the other hand, by using the map $\rho_{[\lambda]}$,
which is a central ingredient in specifying symmetric states with
respect to an action of $SO(3)$, we introduce a reference frame with
respect to which angular momentum will be defined.

Denoting the action of a rotation by an angle $\delta$ around some
given axis $v$ as $R(v,\delta)$, the derivative $\frac{\md}{\md\delta}
R(v,\delta)T_I$ of the action on a spin network state $T_I$ does not
exist. Instead, we are going to define an angular momentum operator
$\hat{L}_v$ by
\begin{equation}\label{AngMom}
 \rho_{[\lambda]}(\hat{L}_vT_I):=-i\hbar\, 
   \left.\left(\frac{\md}{\md\delta}\, \rho_{[\lambda]}(R(v,\delta)T_I)
   \right)\right|_{\delta=0}
\end{equation}
using a different ordering of the pull back $\rho_{[\lambda]}$ and the
derivative in order to render the derivative existing.

To derive the action of $\hat{L}_v$ explicitly, we write a spin
network state as $T_I^{a_1,\ldots,a_n}$ which has $n$ punctures at the
sphere at infinity, each carrying an index $a_p$ in the representation
with label $j_p$ of the intersecting edge (all edges are assumed to be
outgoing at infinity). Note that internal gauge transformations are
frozen at the asymptotic boundary, and therefore each state is a
vector valued function on the space of generalized connections taking
values in the tensor product $j_1\otimes\cdots\otimes j_n$. For the
pull back to invariant connections only the radial part of an
asymptotic holonomy, which is transversal due to the tangle
property, matters and has the form
\[
 h_e=\exp\left(-\int_e\md r A(r)n_r^i(\vartheta,\varphi)\tau_i\right)\,.
\]
After a rotation by an angle $\delta$ around the polar axis $v_3$ of the
coordinates it changes by conjugation to
\[
 h'_e=\exp\left(-\int_e\md r A(r)n_r^i(\vartheta,\varphi+
   \delta)\tau_i\right)=
 \exp(\delta\tau_3) h_e \exp(-\delta\tau_3)\,.
\]
The exponential at the left hand side of $h_e$ corresponds to an
$SU(2)$-transformation at the puncture at infinity, whereas the
exponential at the right hand side corresponds to an inner vertex and
is absorbed due to gauge invariance. We, therefore, have
\[
 \rho_{[\lambda]}\left(R(v_3,\delta)T_I^{a_1,\ldots,a_n}\right)=
 \pi^{j_1}\left(\exp(\delta\tau_3)\right)^{a_1}_{b_1} \cdots
 \pi^{j_n}\left(\exp(\delta\tau_3)\right)^{a_n}_{b_n}
 \rho_{[\lambda]} \left(T_I^{b_1,\ldots,b_n}\right)
\]
because the $p$-th index is in the representation $\pi^{j_p}$.

This expression can easily be differentiated with respect to
$\delta$ yielding by inspection of the general definition
(\ref{AngMom})
\begin{equation}\label{Lthree}
 \hat{L}_3T_I=-i\hbar\,(\pi^{j_1}(\tau_3)\oplus\cdots\oplus
 \pi^{j_n}(\tau_3))\,T_I
\end{equation}
as the third component of the angular momentum operator generating
rotations around the $z$-axis (which has been used as axis for the
polar coordinates). A rotation around an arbitrary axis given by the
direction $v^i$ in $S^2$ leads to a conjugation of asymptotic
holonomies with $\exp(\delta v^i\tau_i)$ and the angular
momentum with respect to this direction is
\begin{equation}\label{L}
 \hat{L}_vT_I=-i\hbar\,(\pi^{j_1}(v^i\tau_i)\oplus\cdots\oplus
 \pi^{j_n}(v^i\tau_i))\,T_I\,.
\end{equation}

Taking the three components $\hat{L}_1$, $\hat{L}_2$, $\hat{L}_3$
corresponding to rotations around Cartesian axes, this immediately
implies the correct commutation relations (which, of course, directly
come from the relation of rotations to internal rotations) of an
angular momentum:
\begin{equation}
 [\hat{L}_i,\hat{L}_j]=-i\hbar\,\epsilon_{ijk}\hat{L}_k\,.
\end{equation}
Furthermore, we can determine the angular momentum spectrum: we just
have to decompose the tensor product of all representations associated
with punctures at infinity into irreducible ones by building
appropriate linear combinations of the components
$T_I^{a_1,\ldots,a_n}$ which transform under one irreducible
subrepresentation. Any component of angular momentum then has
eigenvalues $\hbar m$ where $m=\sum_im_i\in\half\Z$ is given by a sum
over all punctures each of which transforms like a state in the
$j_i$-representation given by $m_i$. Also the absolute value has the
usual eigenvalues $L^{(j)}=\hbar\sqrt{j(j+1)}$, $j\in\half\N_0$ with
eigenstates being given by spin network states. In particular, the
absolute value of the angular momentum operator and the area operator
are simultaneously diagonalizable.  Furthermore, for a given spin
network having a set of spins $\{j_p\}$ labeling its punctures at
infinity, an upper bound for the angular momentum eigenvalues is given
by
\begin{equation}\label{InEq}
 L\leq\hbar\sqrt{{\textstyle\sum_pj_p\left(\sum_pj_p+1\right)}}\,.
\end{equation}

\section{Inequalities between Angular Momentum and Area}

As argued in Ref.\ \cite{Extremal}, an inequality like the last one
has an immediate application to extremal black holes which are
classically defined as saturating the ``no naked singularity''
condition $L\leq(8\pi G)^{-1}A$ between angular momentum and horizon
area of a Kerr black hole. Because the author of Ref.\ \cite{Extremal}
had no angular momentum operator at his disposal, he assumed (without
any concrete justification) the angular momentum eigenvalues to be
given by the spins of a spin network satisfying the inequality
$L\leq\hbar\sum_pj_p$ (although not noted explicitly, it has also been
assumed that angular momentum and area are simultaneously
diagonalizable). Using that, for a given set of punctures, the area
eigenvalues $A=8\pi\gamma\lP^2\sum_p\sqrt{j_p(j_p+1)}$ are bounded
from below by $8\pi\gamma\lP^2\sum_pj_p$, an inequality
$L\leq(8\pi\gamma G)^{-1}A$ for the eigenvalues was derived which
resembles the classical relation. Because the area eigenvalue for a
given $\sum_pj_p$ is minimal if there is only one puncture with spin
$\sum_pj_p$, this one-puncture case was identified with an extremal
black hole.

We will now see what the situation looks like when using our angular
momentum operator. However, there is the important caveat that the
operator (\ref{Lthree}) has been defined at infinity and not, as
needed in Krasnov's argumentation, at the horizon.  Noting that all we
needed for our definition of the angular momentum operator was the
asymptotic form of a connection and the fact that an internal
direction of mapping degree one is fixed at the boundary, we can
immediately apply our discussion of spatial infinity to the case of an
isolated horizon: the isolated horizon boundary conditions also can be
used to fix an internal direction of mapping degree one at the horizon
two-sphere (see, e.g., Refs.\ \cite{IHPhase,IHEntro}).  Although
internal gauge transformations are no longer frozen at an inner
boundary, the boundary punctures of a bulk spin network are the same
as used above because they are coupled to a Chern--Simons state at the
boundary which restores gauge invariance.  All the remaining steps of
the derivation of the angular momentum operator then go through
unaltered.

Now we can compare the eigenvalues of angular momentum and horizon
area.  First, we have a larger upper bound for angular momentum than
assumed in Ref.\ \cite{Extremal}, given by Eq.\ (\ref{InEq}).
However, the lower bound for the area eigenvalues associated with a
set of punctures $\{j_p\}$ can be refined by using the inequality
\begin{eqnarray*}
 \sqrt{(j_1+j_2)(j_1+j_2+1)} & = & \sqrt{j_1^2+j_2^2+2j_1j_2+j_1+j_2}\\
  & \leq & \sqrt{j_1(j_1+1)+ 2\sqrt{j_1(j_1+1)j_2(j_2+1)}+ j_2(j_2+1)}\\
  & = & \sqrt{j_1(j_1+1)}+ \sqrt{j_2(j_2+1)}
\end{eqnarray*}
and induction over the number of punctures. This again leads to an
inequality
\begin{equation}
 \hbar^{-1}L\leq\sqrt{{\textstyle\sum_pj_p\left(\sum_pj_p+1\right)}}\leq 
 (8\pi\gamma\lP^2)^{-1}A
\end{equation}
which is saturated for one-puncture states and differs from the
classical relation only by the factor $\gamma$.  Note that only the
area spectrum is affected by this parameter, whereas the spectrum of
angular momentum is protected against a rescaling by the commutation
relations.

\section{Non-Rotating and Rotating Isolated Horizons}

As another application of the angular momentum operator, we check
whether the non-rotating horizon geometry derived in Ref.\ 
\cite{IHPhase} corresponds to zero angular momentum as seen from the
quantum theory. Recall that non-rotating isolated horizons are defined
classically by suitable boundary conditions which then are used to
select a sector of space-times to be quantized. Classically the
condition of being non-rotating is implemented by requiring the
intrinsic geometry of the horizon to be spherically symmetric. After
quantization, the quantum horizon geometry is described by a
``punctured sphere'' where a spin network in the bulk (outside the
horizon) pierces the horizon in isolated punctures thereby providing
the horizon two-sphere with geometry. Note, incidentally, that the
quantum geometry of a non-rotating horizon is no longer spherically
symmetric which coincides with the observation \cite{SymmRed} that
exactly spherically symmetric states only exist in the sense of
distributions: the quantum horizon geometry of a realistic black hole
cannot be spherically symmetric even if it is non-rotating. Even
finest approximations of a symmetric distribution by ordinary states
will break the symmetry by introducing a discrete set of punctures.

In the framework of Ref.\ \cite{IHPhase} the boundary degrees of
freedom of a non-rotating isolated horizon are described by a
Chern--Simons theory which is ``glued'' to the bulk spin network by a
boundary condition. A bulk state is labeled by the spins
$j=(j_1,\ldots,j_n)\in(\half\N)^n$ of the punctures together with
half-integers $m=(m_1,\ldots,m_n)\in(\half\Z)^n$ which determine a
state in the product of all representations labeling the punctures,
i.e., they are subject to the conditions
$m_i\in\{-j_i,-j_i+1,\ldots,j_i\}$. Given such a combination of spin
labels of the bulk state, a permissible boundary state of the
Chern--Simons theory is determined by numbers
$a=(a_1,\ldots,a_n)\in\Z_k^n$ fulfilling $a_i\equiv -2m_i\mod k$ and
$\sum_ia_i\equiv 0\mod k$ where $k=\frac{a_0}{4\pi\gamma\lP^2}$ is the
level of the Chern--Simons theory and related to the classical horizon
area $a_0$ (a prescribed parameter). The first condition on $a$
describes how the bulk states labeled by $(j,m)$ are glued to a
Chern--Simons state labeled by $a$, and the second condition arises
from gauge invariance in Chern--Simons theory and can be interpreted
as saying that the sum of all deficit angles introduced by the
punctures vanishes modulo $4\pi$ \cite{IHEntro}.

Using our angular momentum operator, we can see another interpretation
of the condition $\sum_ia_i\equiv0$ if we first transfer it via the
gluing conditions to the bulk labels resulting in $\sum_im_i=0$. Here
we ignored the fact that the sum of the $a_i$ vanishes only modulo $k$
which can be seen only in the Chern--Simons boundary theory. However,
for any combination of labels $m_i$ subject to the condition
$\sum_im_i=0$ we can find a permissible set of labels $a_i$. The
condition $\sum_im_i=0$ in turn says that the bulk state can be
associated with the trivial representation in the tensor product of
all puncture representations and so has vanishing angular momentum
as measured with the operator (\ref{Lthree}). Thus, the condition
$\sum_ia_i\equiv 0$ on the non-rotating quantum horizon state can be
interpreted naturally as saying that the classical property of being
non-rotating is preserved after quantization.

Our interpretation of the boundary states also indicates how the
quantum geometry of non-rotating horizons could be generalized to
rotating ones. In fact, on a bulk state we just have to replace the
condition $\sum_im_i=0$ by a condition $\sum_im_i=l_0\not=0$ where
$l_0$ is a given (analogously to the classical horizon area $a_0$)
value of the angular momentum. However, this argument tells nothing
about how to generalize the Chern--Simons boundary theory to rotating
isolated horizons which could only be derived by a Hamiltonian
analysis using more general boundary conditions.

Assuming that the generalization to rotating horizons is correct we
can check whether the successful calculation of the entropy of
non-rotating black holes (possibly charged and non-extremal)
\cite{IHEntro} remains valid for rotating black holes. In fact, it is
quite easy to see that this is the case for rotating black holes not
too close to extremality ($(8\pi\gamma\lP^2)^{-1}a_0\gg l_0$) using
the methods of Ref.\ \cite{IHEntro}: First, a lower bound for the
number of states with prescribed area around $a_0$ and angular
momentum around $\hbar l_0$ can be derived by using configurations $j$
with $j_1=\cdots=j_{n-1}=\half$ and $j_n=l_0$ together with
$m_1=\cdots=m_{n-1}=\pm\half$ and $m_n=l_0$. The condition
$\sum_{i=1}^nm_i=l_0$ then is equivalent to $\sum_{i=1}^{n-1}m_i=0$
and results in a number
\[
  N_{\mathrm{bh}}\geq \frac{2^{n-\frac{1}{2}}}{\sqrt{\pi (n-1)}}
\]
of states (just replace $n$ with $n-1$ in the corresponding formula of
Ref.\ \cite{IHEntro}). Taking the logarithm and using the condition of
not being near-extremal we obtain the same lower bound for the entropy
\[
  S_{{\mathrm{bh}}}\geq\frac{\log
    2}{4\pi\sqrt{3}\gamma\lP^2}\,a_0-o(a_0)
\]
as in Ref.\ \cite{IHEntro}. Because the derivation of an upper bound
for the entropy in Ref.\ \cite{IHEntro} did not make use of the
condition $\sum_ia_i\equiv0$, we can immediately transfer it to the
rotating case and arrive at our result that the recent calculations of
black hole entropy in loop quantum gravity remain valid without
changes also for rotating (possibly charged but far from extremal)
black holes:
\begin{equation}
 S_{{\mathrm{bh}}}=\frac{\log
    2}{4\pi\sqrt{3}\gamma\lP^2}\,a_0+o(a_0)\,.
\end{equation}
In particular, the Bekenstein--Hawking formula can be obtained with
the correct numerical factor not
only for charged but also for rotating black holes by fixing the
Immirzi parameter to be
\begin{equation}
 \gamma_0=\frac{\log 2}{\pi\sqrt{3}}\,.
\end{equation}

For near-extremal rotating black holes, however, the entropy is
reduced. In the extremal case we have
$a_0=8\pi\gamma\lP^2\sqrt{l_0(l_0+1)}$ and according to Ref.\ 
\cite{Extremal} the boundary state has a single $l_0$-puncture with at
most $2l_0+1$ values for $m$. This results in
$S_{\mathrm{extr}}\leq\log((4\pi\gamma\lP^2)^{-1}a_0)$ being at most
logarithmic in the area.

To conclude, we note that as demonstrated here the methods developed in
Ref.\ \cite{SymmRed} are not only applicable in the study of reduced
models but also provide tools for direct applications to the full
theory.

\section*{Acknowledgements}

The author thanks H Kastrup for discussions and the
DFG-Graduierten-Kolleg ``Starke und elektroschwache Wechselwirkung bei
hohen Energien'' for a PhD fellowship.

\end{document}